\documentclass[11pt]{article}
\usepackage{moriond}
\usepackage{xspace}
\usepackage{wrapfig}
\usepackage{amssymb}
\usepackage{amsmath}
\usepackage{url}

\usepackage[numbers,sort&compress]{natbib}
\usepackage{natbibspacing}
\usepackage{hypernat}
\newcommand\allFontSize{\small}

\newenvironment{myquote}
               {\list{}{\leftmargin0cm}%
                \item\relax}
               {\endlist}

\usepackage[bf]{caption}

\newcommand\detailsSize{\allFontSize}
{\begin{myquote}\vspace{-0.2cm}\detailsSize}{\end{myquote}\vspace{-0.2cm}}

{\setlength{\gfitterboxwidth}{\textwidth}\addtolength{\gfitterboxwidth}{-2.3\fboxsep}%
\begin{Sbox}\begin{tabular*}{\gfitterboxwidth}{>{\rule[-0.5ex]{.0ex}{3.0ex}\tt\small}p{3cm}>{\tt\small}p{3.5cm}>{\small}p{6.7cm}}
&&\\[-0.7cm]
\rm\small Option  & \rm\small Values    & \rm\small Description\\[0.1cm]\hline
&&\\[-0.45cm]}%
{\\[-0.1cm]\end{tabular*}\end{Sbox}\vspace{1ex} \fbox{\TheSbox}}

{\setlength{\gfitterboxwidth}{\textwidth}\addtolength{\gfitterboxwidth}{-2.3\fboxsep}%
\begin{Sbox}\begin{tabular*}{\gfitterboxwidth}{>{\rule[-0.5ex]{.0ex}{3.5ex}\tt\small}p{3.0cm}>{\tt\small}p{3.0cm}>{\small}p{7.2cm}}
&&\\[-0.7cm]
\rm\small Option  & \rm\small Values    & \rm\small Description\\[0.1cm]\hline
&&\\[-0.45cm]}%
{\\[-0.1cm]\end{tabular*}\end{Sbox}\vspace{1ex} \fbox{\TheSbox}}

{\setlength{\gfitterboxwidth}{\textwidth}\addtolength{\gfitterboxwidth}{-2.3\fboxsep}%
\begin{Sbox}\begin{tabular*}{\gfitterboxwidth}{>{\rule[-0.5ex]{.0ex}{3.0ex}\tt\small}p{2.5cm}>{\tt\small}p{3.5cm}>{\small}p{7.2cm}}
&&\\[-0.7cm]
\rm\small Option  & \rm\small Values    & \rm\small Description\\[0.1cm]\hline
&&\\[-0.45cm]}%
{\\[-0.1cm]\end{tabular*}\end{Sbox}\vspace{1ex} \fbox{\TheSbox}}

{\setlength{\gfitterboxwidth}{\textwidth}\addtolength{\gfitterboxwidth}{-2\fboxsep}%
\begin{Sbox}\begin{tabular*}{\gfitterboxwidth}{>{\rule[-0.5ex]{.0ex}{3.0ex}\tt\small}p{4.5cm}>{\tt\small}p{3.2cm}>{\small}p{5.5cm}}
&&\\[-0.7cm]
\rm\small Option  & \rm\small Values    & \rm\small Description\\[0.1cm]\hline
&&\\[-0.45cm]}%
{\\[-0.1cm]\end{tabular*}\end{Sbox}\vspace{1ex} \fbox{\TheSbox}}

{\setlength{\gfitterboxwidth}{\textwidth}\addtolength{\gfitterboxwidth}{-2\fboxsep}%
\begin{Sbox}\begin{tabular*}{\gfitterboxwidth}{>{\rule[-0.5ex]{.0ex}{3.0ex}\tt\small}p{4.0cm}>{\tt\small}p{3.0cm}>{\small}p{6.2cm}}
&&\\[-0.7cm]
\rm\small Option  & \rm\small Values    & \rm\small Description\\[0.1cm]\hline
&&\\[-0.45cm]}%
{\\[-0.1cm]\end{tabular*}\end{Sbox}\vspace{1ex} \fbox{\TheSbox}}

\newcommand\Gfitter{Gfitter\xspace}

\newcommand{\bbar}  {\ensuremath{\overline b}\xspace}

\mathchardef\Upsilon="7107
\def\Y#1S{\ensuremath{\Upsilon{(#1S)}}\xspace}

\newcommand{\mt}{\ensuremath{m_{t}}\xspace}
\newcommand{\as}{\ensuremath{\alpha_{\scriptscriptstyle S}}\xspace}

\newcommand{\asZ}{\ensuremath{\as(M_Z^2)}\xspace}

\renewcommand\l{\ell}

\newcommand{\Afbz}[1]{{\ensuremath{A_{\rm\scriptscriptstyle FB}^{0,#1}}}\xspace}

\newcommand{\GF}{{\ensuremath{G_{\scriptscriptstyle F}}}\xspace}

\newcommand{\pZ}{\ensuremath{Z}\xspace}
\newcommand{\pW}{\ensuremath{W}\xspace}

\newcommand{\Kbar    }{\kern 0.2em\overline{\kern -0.2em K}{}\xspace}

\newcommand{\Kz      }{\ensuremath{K^0}\xspace}
\newcommand{\Kzb     }{\ensuremath{\Kbar^0}\xspace}
\newcommand{\KzKzb   }{\ensuremath{\Kz \kern -0.16em \Kzb}\xspace}
\newcommand{\Kp      }{\ensuremath{K^+}\xspace}
\newcommand{\Km      }{\ensuremath{K^-}\xspace}

\newcommand{\KpKm    }{\ensuremath{\Kp \kern -0.16em \Km}\xspace}

\newcommand\Dbar    {\kern 0.18em\overline{\kern -0.18em D}{}\xspace}

\newcommand\Bbar    {\kern 0.18em\overline{\kern -0.18em B}{}\xspace}

\newcommand\Bz      {\ensuremath{B^0}\xspace}

\newcommand\Bzb     {\ensuremath{\Bbar^0}\xspace}
\newcommand\Bu      {\ensuremath{B^+}\xspace}
\newcommand\Bub     {\ensuremath{B^-}\xspace}

\newcommand\BpBm    {\ensuremath{\Bu {\kern -0.16em \Bub}}\xspace}
\newcommand\Bs      {\ensuremath{B^0_{s}}\xspace}
\newcommand\Bsb     {\ensuremath{\Bbar^0_{s}}\xspace}

\newcommand\BzBzb   {\ensuremath{\Bz {\kern -0.16em \Bzb}}\xspace}
\newcommand\BszBszb {\ensuremath{\Bs {\kern -0.16em \Bsb}}\xspace}

\newcommand\deltatheo{\ensuremath{\delta_{\rm th}}\xspace}

\newcommand\Rfit{{\em R}fit\xspace}

\newcommand{\tev}{\ensuremath{\mathrm{Te\kern -0.1em V}}\xspace}
\newcommand{\gev}{\ensuremath{\mathrm{Ge\kern -0.1em V}}\xspace}
\newcommand{\mev}{\ensuremath{\mathrm{Me\kern -0.1em V}}\xspace}
\newcommand{\kev}{\ensuremath{\mathrm{ke\kern -0.1em V}}\xspace}
\newcommand{\ev}{\ensuremath{\mathrm{e\kern -0.1em V}}\xspace}
\newcommand{\gevc}{\ensuremath{{\mathrm{Ge\kern -0.1em V\!/}c}}\xspace}
\newcommand{\mevc}{\ensuremath{{\mathrm{Me\kern -0.1em V\!/}c}}\xspace}
\newcommand{\gevcc}{\ensuremath{{\mathrm{Ge\kern -0.1em V\!/}c^2}}\xspace}
\newcommand{\mevcc}{\ensuremath{{\mathrm{Me\kern -0.1em V\!/}c^2}}\xspace}

\newcommand{\bei}{\begin{itemize}}
\newcommand{\eei}{\end{itemize}}
\newcommand{\beq}{\begin{equation}}
\newcommand{\eeq}{\end{equation}}
\newcommand{\beqn}{\begin{eqnarray}}
\newcommand{\eeqn}{\end{eqnarray}}
\newcommand{\beqns}{\begin{eqnarray*}}
\newcommand{\eeqns}{\end{eqnarray*}}
\newcommand{\bitm}{\begin{itemize}}
\newcommand{\eitm}{\end{itemize}}

\newcommand{\dahadZf}{\ensuremath{\Delta\alpha_{\rm had}^{(5)}(M_Z^2)}\xspace}
\newcommand{\dalphaHadMZ}{\ensuremath{\Delta\alpha_{\rm had}^{(5)}(M_Z^2)}\xspace}
\newcommand{\dalphaHads}{\ensuremath{\Delta\alpha_{\rm had}^{(5)}(s)}\xspace}

\newcommand{\MH}{\ensuremath{M_{H}}\xspace}
\newcommand{\MW}{\ensuremath{M_{W}}\xspace}
\newcommand{\MZ}{\ensuremath{M_{Z}}\xspace}
\newcommand{\Rzb}{\ensuremath{R^0_b}\xspace}

\newcommand{\STU}{$S,\,T,\,U$\xspace}
\def\@citex[#1]#2{\if@filesw\immediate\write\@auxout{\string\citation{#2}}\fi
  \@tempcnta\z@\@tempcntb\m@ne\def\@citea{}\@cite{\@for\@citeb:=#2\do
    {\@ifundefined
       {b@\@citeb}{\@citeo\@tempcntb\m@ne\@citea
        \def\@citea{,\penalty\@m\ }{\bf ?}\@warning
       {Citation `\@citeb' on page \thepage \space undefined}}%
    {\setbox\z@\hbox{\global\@tempcntc0\csname b@\@citeb\endcsname\relax}%
     \ifnum\@tempcntc=\z@ \@citeo\@tempcntb\m@ne
       \@citea\def\@citea{,\penalty\@m}
       \hbox{\csname b@\@citeb\endcsname}%
     \else
      \advance\@tempcntb\@ne
      \ifnum\@tempcntb=\@tempcntc
      \else\advance\@tempcntb\m@ne\@citeo
      \@tempcnta\@tempcntc\@tempcntb\@tempcntc\fi\fi}}\@citeo}{#1}}

\def\@citeo{\ifnum\@tempcnta>\@tempcntb\else\@citea
  \def\@citea{,\penalty\@m}%
  \ifnum\@tempcnta=\@tempcntb\the\@tempcnta\else
   {\advance\@tempcnta\@ne\ifnum\@tempcnta=\@tempcntb \else
\def\@citea{--}\fi
    \advance\@tempcnta\m@ne\the\@tempcnta\@citea\the\@tempcntb}\fi\fi}

\newcommand\mini{{\rm min}}

\newcommand\ChiMin{\ensuremath{\chi^2_{\mini}}\xspace}

\newcommand{\seffsf}[1]{\ensuremath{\sin\!^2\theta^{#1}_{{\rm eff}}}\xspace}
\newcommand{\sinleff}{\seffsf{\ell}}

\newcommand{\WMass}     	{\ensuremath{80.385\pm0.015\:\gev}}
\newcommand{\WWidth}     	{\ensuremath{2.085\pm0.042\:\gev}}

\newcommand{\TopMassInd}        {\ensuremath{175.8^{\:+2.7}_{\:-2.4}\:\gev}}

\newcommand{\SParam}     	{\ensuremath{0.03\pm 0.10}\xspace}
\newcommand{\TParam}     	{\ensuremath{0.05\pm 0.12}\xspace}
\newcommand{\UParam}     	{\ensuremath{0.03\pm 0.10}\xspace}

\newcommand{\STParamCor}	{\ensuremath{+0.89}\xspace}
\newcommand{\SUParamCor}	{\ensuremath{-0.54}\xspace}
\newcommand{\TUParamCor}	{\ensuremath{-0.83}\xspace}

\newcommand{\SParamNU}     	{\ensuremath{0.05\pm 0.09}\xspace}
\newcommand{\TParamNU}     	{\ensuremath{0.08\pm 0.07}\xspace}
\newcommand{\STParamCorNo}     	{\ensuremath{+0.91}\xspace}

\newcommand{\Photo}{}

\begin{document}


\vspace*{4cm}
\title{The global electroweak Standard Model fit after the Higgs discovery}

\author{ M. Baak$^{a}$, R. Kogler$^{b}$ (for the Gfitter group)}
\address{$^{a}$CERN, Geneva, Switzerland \\ $^{b}$Institut f\"ur Experimentalphysik, Universit\"at Hamburg, Germany}

\maketitle\abstracts{
We present an update of the global Standard Model (SM) fit to electroweak precision data under the assumption that  the new particle discovered at the LHC is the SM Higgs boson. In this scenario all parameters entering the calculations of electroweak precision observalbes are known, allowing, for the first time, to over-constrain the SM at the electroweak scale and assert its validity.
Within the SM the \pW boson mass and the effective weak mixing angle can be accurately predicted from the global fit. The results are compatible with, and exceed in precision, the direct measurements. An updated determination of the $S$, $T$ and $U$ parameters, which parametrize the oblique vacuum corrections, is given. The obtained values show good consistency with the SM expectation and no direct signs of new physics are seen.
We conclude with an outlook to the global electroweak fit for a future $e^+e^-$ collider.
}

\section{Introduction}
The electroweak fit of the Standard Model (SM) has a long tradition in particle physics. It relies on the predictability of the SM, where all electroweak observables can be expressed as functions of five parameters.
A tremendous amount of pioneering work from the theoretical community in the calculation of radiative corrections, as well as from the experimental community in the measurement of electroweak precision observables, led to a correct prediction~\cite{LEPEWWG:1995ac} of the top-quark mass \mt before its actual discovery 
in 1995~\cite{Abe:1995hr, Abachi:1995iq}\,\footnote{ 
The observed top quark mass was $176\pm8$(stat)$\pm10$(sys.)$\:\gev$ and $199^{+19}_{-21}$(stat.)$\pm22$(sys.)$\:\gev$ as measured by the CDF and D0 collaborations. This is in very good agreement with the SM prediction of $\mt=178^{+19}_{-22}$~\gev~\cite{Campagnari:1996ai}, as obtained from electroweak precision data.
}.
This success has given confidence in the calculations of radiative corrections, including loop-contributions from the Higgs boson. 

The discovery of the top quark left the Higgs boson mass \MH as the last free parameter of the SM without experimental constraints. 
The focus of the electroweak fits shifted to precisely predicting \MH from electroweak precision observables (EWPO). However, while the loop corrections to the \pW- and \pZ-propagators involving the top quark lead to an approximate quadratic dependence, \MH enters only logarithmically in the calculation of electroweak observables, leading to weaker constraints on \MH than on \mt. Nevertheless, improvements in theoretical and experimental precision, especially on \mt, \MW and the hadronic contribution to the running of the electromagnetic coupling for the five light quarks \dalphaHads, led to a rather precise SM prediction of $\MH=96^{+31}_{-24}$~\gev~\cite{Baak:2011ze}.

This was the status of the electroweak fit at the discovery of the new Higgs-like boson reported by the ATLAS and CMS collaborations~\cite{ATLAS:2012gk, CMS:2012gu}
The predicted value is in very good agreement with its measured mass of $\sim\!126\:\gev$.
Most recent measurements and analyses of the properties of this newly discovered boson, as presented at this conference, 
show good consistency with the assumption that the new particle is indeed the SM Higgs boson~\cite{Dumont:2013mba}.

Supposing that the new particle is the SM Higgs boson, all parameters entering electroweak precision observables are known, allowing a full assessment of the consistency of the SM at the electroweak scale~\cite{Eberhardt:2012gv}.
We interpret the new particle as the SM Higgs boson and present an update of the SM electroweak fit using the \Gfitter framework~\cite{Flacher:2008zq}. 
The \Gfitter results shown here have been recently published elsewhere~\cite{Baak:2012kk}.

A detailed study of the implications of the value of \MH as input to the electroweak fit shows an improvement in precision on the predictions of \MW and the effective weak mixing angle \sinleff. We also report updated constraints on the oblique parameters \STU, which parametrize possible contributions to oblique vacuum corrections from physics beyond the SM (BSM) and hence allow to constrain new models through EWPO. We also use the projected experimental uncertainties from a future $e^{+}e^{-}$ facility to derive the expected precision of SM predictions for electroweak observables.
  
\section{The global electroweak fit with \Gfitter}

A detailed description of the calculations and experimental input used in the electroweak fit is given elsewhere~\cite{Baak:2012kk} and only the most important features are given here. The mass of the \pW boson and the effective weak mixing angle are calculated at two-loop order, including leading terms beyond the two loop calculation~\cite{Awramik:2003rn,Awramik:2006uz,Awramik:2004ge}. 
We use the full $\mathcal{O}(\as^4)$ calculation~\cite{Baikov:2008jh, Baikov:2012er} of the QCD Adler function, which stabilizes the perturbative QCD expansion in the calculation of the \pZ-boson width. 
An improved prediction of \Rzb, the hadronic partial decay width of $\pZ \to b \bbar$, is used, which includes the complete calculation of fermionic two-loop corrections~\cite{Freitas:2012sy}. 
The calculation of the vector and axial-vector couplings, $g_A^f$ and $g_V^f$, used in the calculation of the partial and total widths of the \pZ and \pW bosons, relies on accurate parametrizations~\cite{Hagiwara:1994pw,Hagiwara:1998yc,Cho:1999km,Cho:2011rk}.
Theoretical uncertainties are implemented using the \Rfit scheme, which 
%
corresponds to a linear addition of theoretical and experimental uncertainties. 
The two uncertainties considered are due to missing higher orders in the calculations of \MW and \sinleff and have been estimated to be $\deltatheo\MW = 4\:\mev$~\cite{Awramik:2003rn} and $\deltatheo\sinleff = 4.7\cdot10^{-5}$~\cite{Awramik:2006uz}.

The experimental input used in the fit include the electroweak precision data measured at the \pZ-pole together with their correlations~\cite{LEPEWWG:2005ema}. For the mass and width of the \pW boson the latest world average is used, $\MW = \WMass$ and $\Gamma_W= \WWidth$, obtained from measurements by the LEP and Tevatron experiments~\cite{TevatronElectroweakWorkingGroup:2012gb}. 
While the leptonic contribution to the running of the electromagnetic coupling strength can be calculated with very high precision, the value of the hadronic contribution is obtained from a fit to experimental data supplemented by perturbative calculations, $\dahadZf=(2757\pm10)\cdot10^{-5}$~\cite{Davier:2010nc}. 

For the mass of the top quark we use the average from the direct measurements by the Tevatron experiments $\mt = 173.18 \pm 0.94\:\gev$~\cite{Aaltonen:2012ra}. Biases in the measurement of \mt due to a mis-modeling of non-perturbative color-reconnection effects in the fragmentation process, initial and final state radiation and kinematics of the $b$-quark, have been studied by the CMS collaboration~\cite{CMS:2012ixa}. 
With the current precision no significant deviation is observed between the measured values and the predictions using different models. 
An additional ambiguity in the interpretation of \mt originates from the top's finite decay width,
with additional uncertainty which is difficult to estimate quantitatively. 
The effect of an additional theoretical uncertainty of $0.5\:\gev$ on \mt has been studied, and the fit shows only a slight deterioration in precision.  
This uncertainty is not included in the standard fit setup.

A na\"{\i}ve combination of the measured values of \MH from the ATLAS and CMS experiments as reported in \cite{ATLAS:2012gk, CMS:2012gu}, gives $M_H=125.7 \pm 0.4\:\gev$, where the systematic uncertainties are treated as fully uncorrelated. Treating them as fully correlated only changes the uncertainty to $0.5\:\gev$, with a negligible effect on the fit result due to the weak dependence on \MH. 
%

\section{The SM fit}

\begin{wrapfigure}{R}{0pt}
\includegraphics[width=0.4\textwidth]{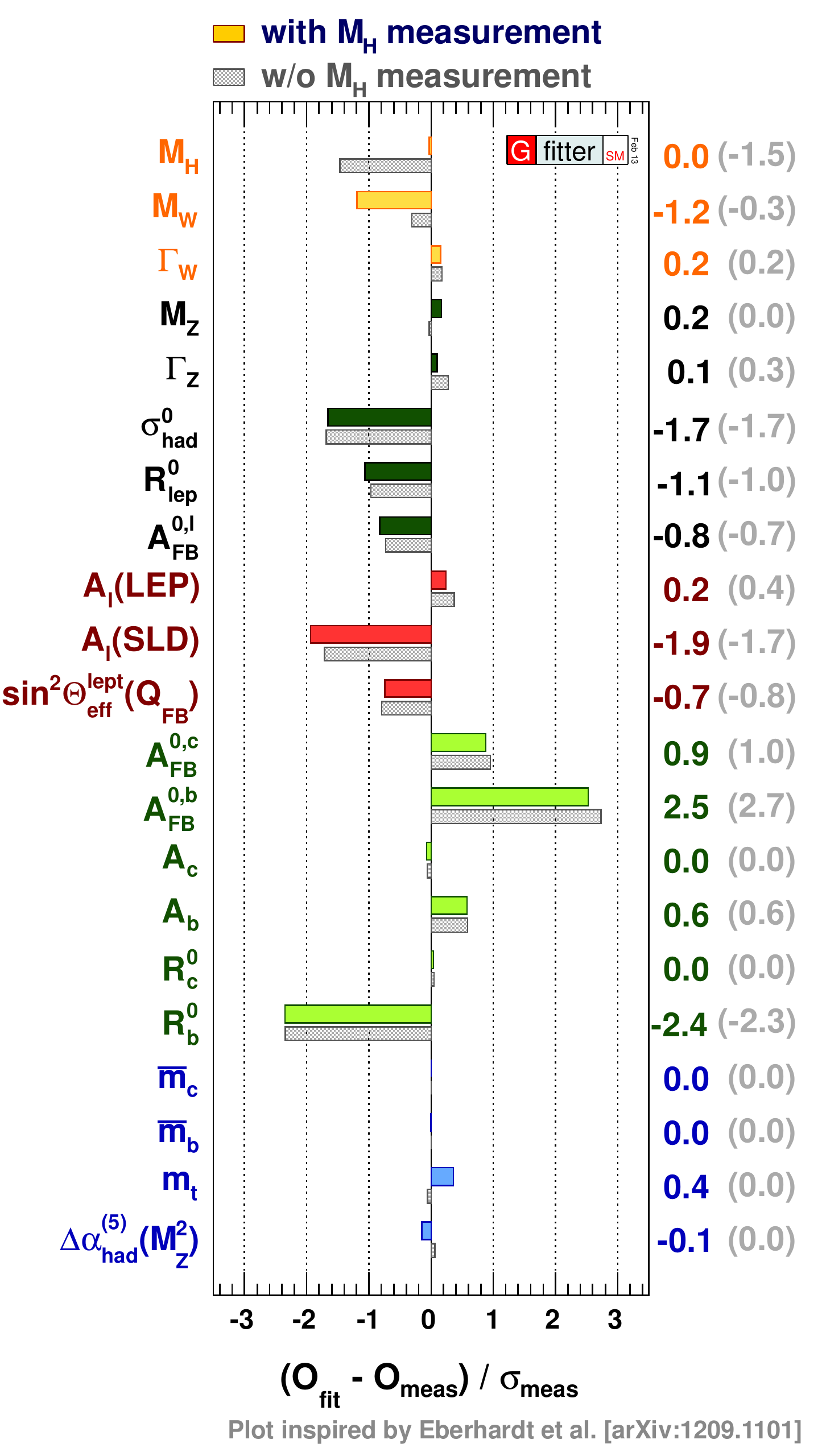}
\caption{Differences between the SM prediction and the measured parameter, in units of the uncertainty for the fit including \MH (color) and without \MH (grey). \label{fig:pulls}}
\end{wrapfigure}
The SM electroweak fit is performed in three scenarios~\cite{Baak:2012kk}. In the first scenario all input parameters are used, allowing to test the validity of the SM. The results are compared to the second scenario, where the fit is performed without the inclusion of \MH to assess the effect of knowing \MH in the electroweak fit. In the third scenario individual observables are removed one by one from the fit which allows for an indirect determination of these with an accurate uncertainty calculation. 

The SM fit including all input data converges with a minimum value of the test statistics of $\ChiMin= 21.8$, obtained for $14$ degrees of freedom. Calculating the na\"{\i}ve $p$-value gives ${\rm Prob}(21.8,14)= 0.08$. The smallness of the $p$-value with respect to previous results~\cite{Baak:2011ze} is not due to the inclusion of \MH, but rather due to the new calculation of \Rzb which has a very small dependence on \MH, as described below. 

Performing the fit without \MH as input parameter, the fit converges at a minimum of $\ChiMin= 20.3$ for $13$ degrees of freedom, corresponding to a $p$-value of $0.09$. In this case the fit converges for a value of $\MH=94^{+25}_{-22}\:\gev$, in good agreement with the direct measurement. 

The result of the fit is shown in Fig.~\ref{fig:pulls} in terms of the pull value, which is defined as deviation between the SM prediction and the measured parameter in units of the measurement uncertainty. The fit results are shown for both scenarios, including the \MH measurement (colored bars) and without \MH (grey bars), where in general the result of the fit does not change significantly between the two scenarios.  Very small pull values, as for example observed for the light quark masses but also for \MH, indicate that the input accuracy exceeds the fit requirements. 
No single pull value exceeds $3\sigma$, showing an overall satisfying consistency of the SM.

The largest deviations between the SM prediction and the measurements are observed in the b-sector. Both observables directly sensitive to $\pZ \to b \bbar$, the forward-backward asymmetry \Afbz{b} and the partial width \Rzb, show large deviations of $2.5\sigma$ and $-2.4\sigma$, respectively. While the effect in \Afbz{b} has been known for a long time, the large deviation in \Rzb is new, owing to the improved two-loop calculation which exhibits an unexpected large negative correction~\cite{Freitas:2012sy}. 
Using the one-loop result for \Rzb only, the pull value is $-0.8\sigma$. 
Both parameters show only very little dependence on the inclusion of \MH, with deviations of $2.7\sigma$ and $-2.3\sigma$ in the fit scenario without including \MH.\,\footnote{ 
It is intriguing to observe that an increase of the right-handed coupling of the $\pZ \to b \bbar$ vertex of 25\%, while leaving the left-handed coupling unchanged, can resolve both deviations. 
}

\begin{figure}
\includegraphics[width=0.47\textwidth]{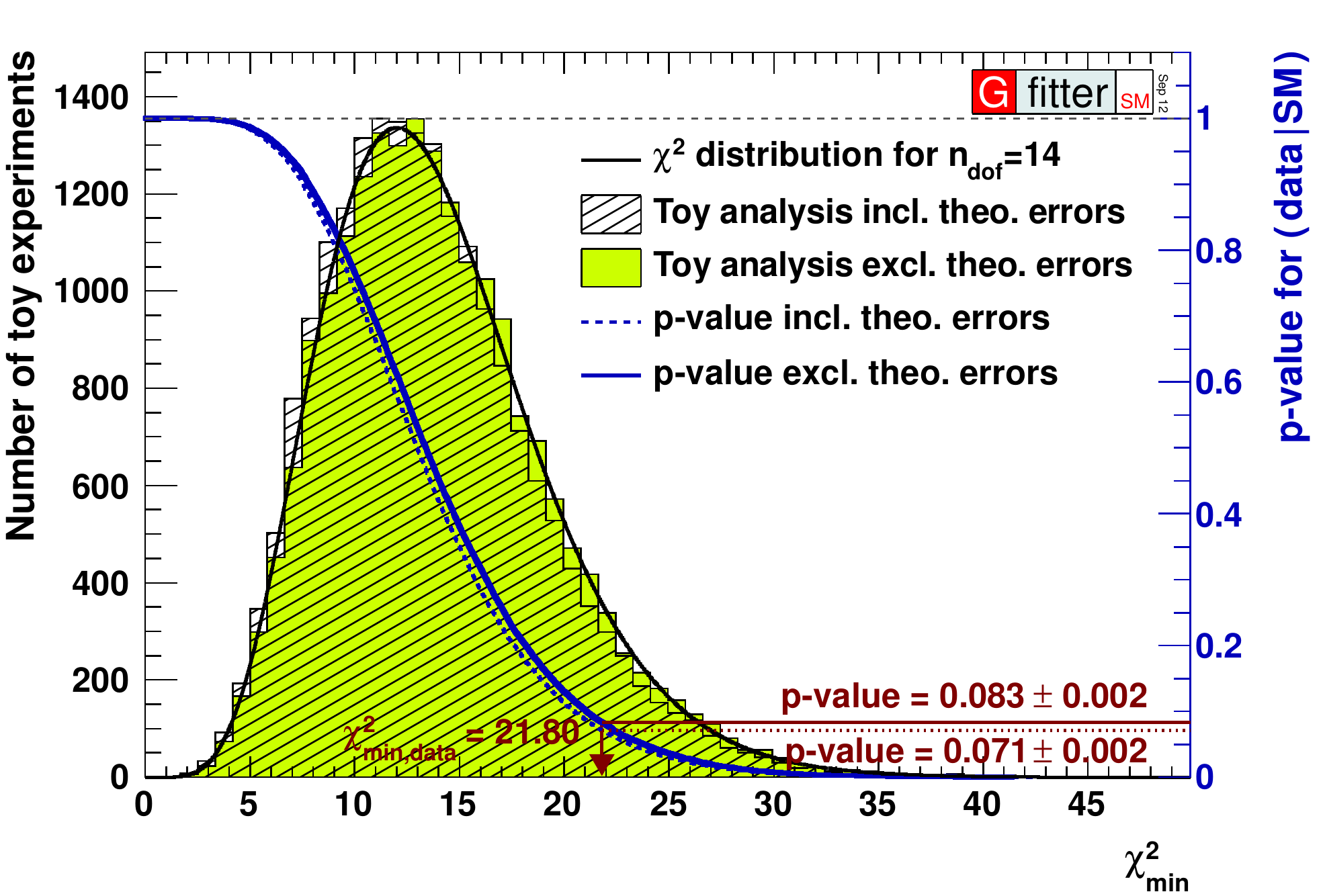} \put(-120, 70){\small (a)} \hspace{0.2cm}
\includegraphics[width=0.5\textwidth]{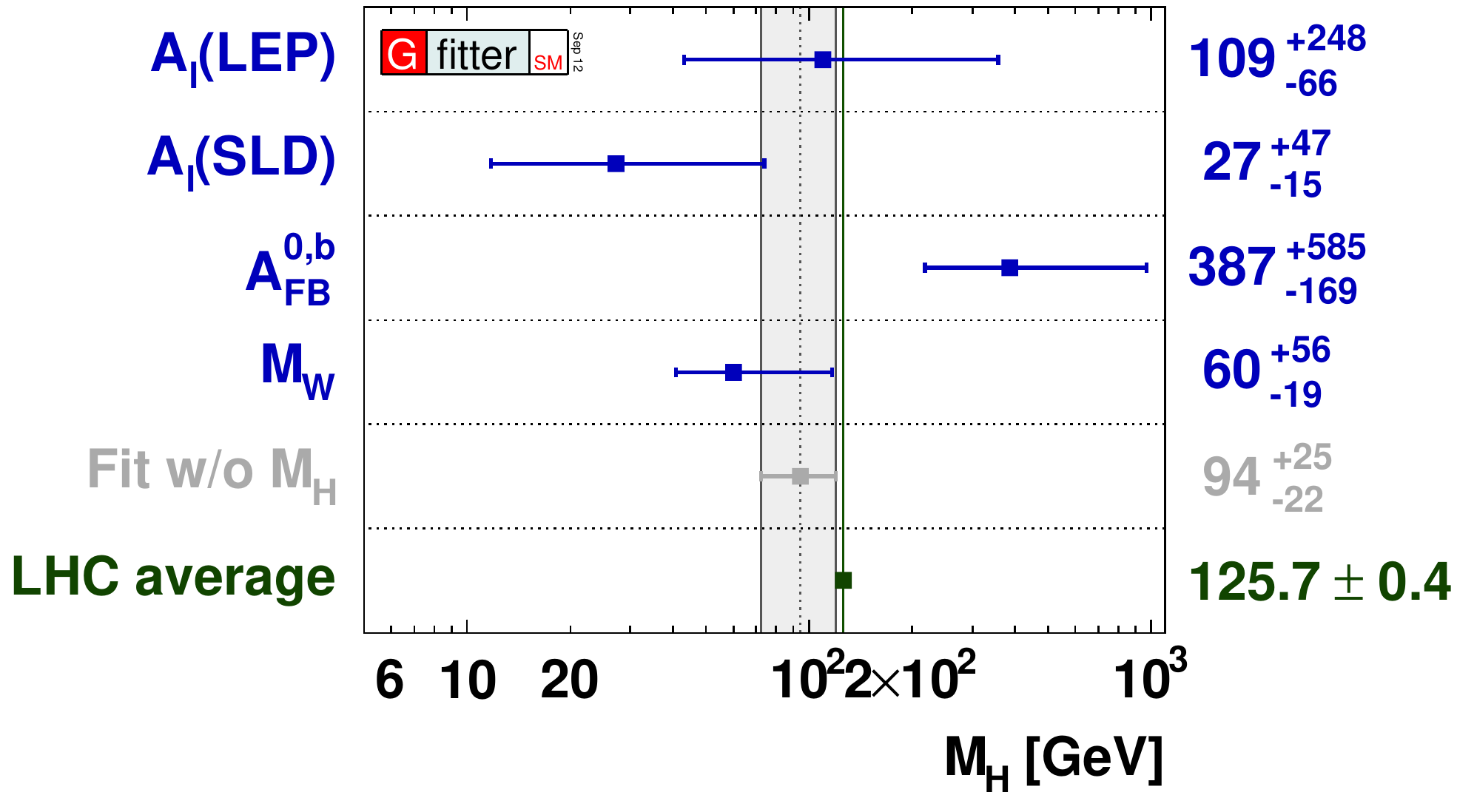} \put(-170, 135){\small (b)} 
\caption{
Distribution of the $\ChiMin$ value obtained from pseudo Monte Carlo simulations (a). Shown are distributions obtained by including (hatched) and excluding (green) the theory uncertainties \deltatheo, compared with the idealized $\chi^2$ distribution assuming Gaussian distributed errors with 14 degrees of freedom. The arrows indicate the $\ChiMin$ value obtained from the fit to data. The result from a determination of \MH using only the given observable is shown in (b). 
\label{fig:toys_mainobs}}
\end{figure}

In order to assess the validity of the fit we use Monte Carlo simulation to generate pseudo experiments. For each simulation we generate SM parameters according to Gaussian distributed values around their expected values with standard deviations equal to the full experimental uncertainty.
The obtained $\ChiMin$ distribution for all toy datasets is shown in Fig.~\ref{fig:toys_mainobs}(a). Good agreement between the MC simulation and the idealized distribution for 14 degrees of freedom is found. The result from the fit to data is indicated as red arrow and the obtained $p$-value is consistent between the MC simulation and the idealized $\chi^2$ distribution.
The influence of the theoretical uncertainties on the $p$-value of the full SM fit amounts to about 0.01.

Except for the value of \MH itself, the largest change in the result due to the inclusion of \MH is the prediction of \MW. 
For this observable the pull value changes from $-0.3$ to $-1.2$, due to the small value of \MH preferred by the \MW measurement. 
This effect is shown in Fig.~\ref{fig:toys_mainobs}(b), where indirect determinations of \MH are displayed, obtained by removing all sensitive observables from the fit except the given one. For comparison, also the indirect fit result using all input parameters except for \MH (grey band) and the direct measurement (green line) are shown. The values obtained from the fit including the measurements of the leptonic asymmetries $A_{\ell}$, as measured by the LEP and SLD collaborations, and \MW show good agreement. 
%
%
The value of \MH obtained from the hadronic forward-backward asymmetry \Afbz{b} shows a tendency towards large values of \MH, with a discrepancy of $2.5\sigma$.

\section{Predictions for key observables}

The inclusion of \MH in the fit results in a large improvement in precision for the indirect determination of several SM parameters. Without the inclusion of \MH, the indirect determination of the top mass gives $\mt = 171.5^{+8.9}_{-5.3}\:\gev$. Including the knowledge about \MH, the fit value obtained is 
\beqn
\mt = \TopMassInd,
\label{eq:mt}
\eeqn
where the uncertainty is reduced by a factor of $2$--$3$. The value of \mt agrees well with the direct determination~\cite{Aaltonen:2012ra} and the cross-section based determination under the assumption that there is no new physics contributing to the cross section measurement~\cite{Alekhin:2012sx}.

\begin{figure}
\begin{center}
\includegraphics[width=0.5\textwidth]{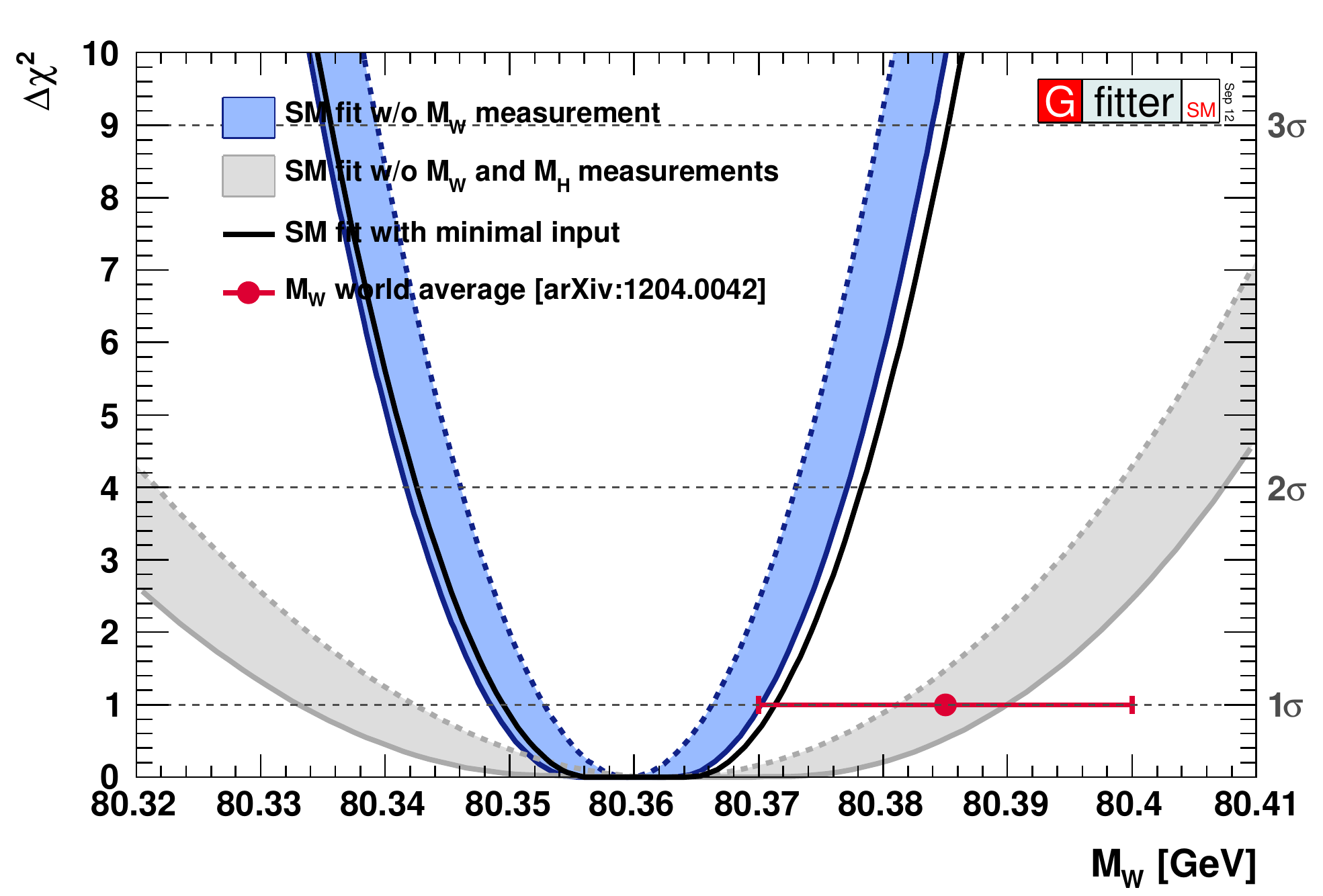} \put(-190, 88){\small (a)}
\includegraphics[width=0.5\textwidth]{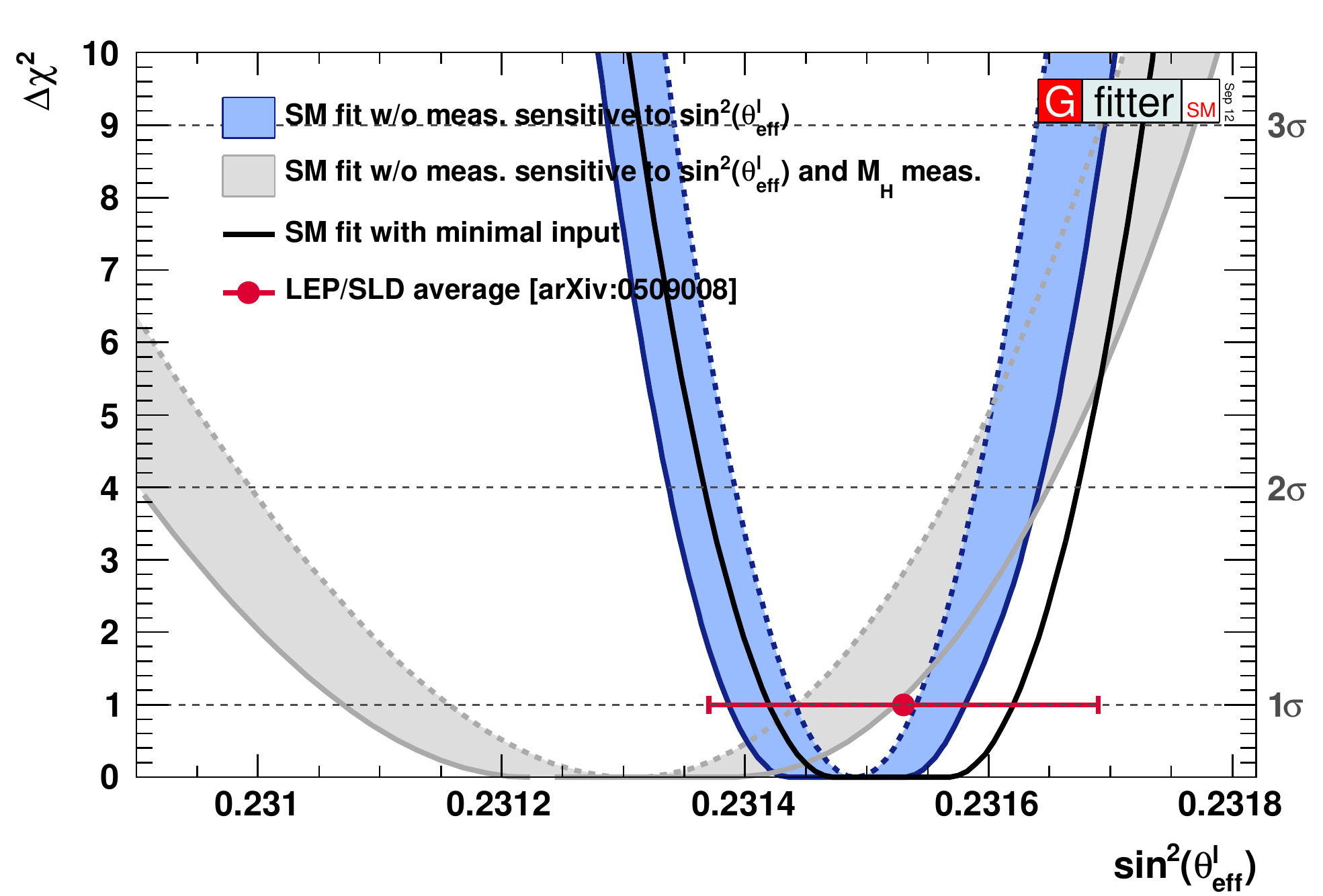} \put(-190, 88){\small (b)} 
\end{center}
\caption{
$\Delta\chi^2$ profiles for the indirect determination of \MW (a) and \sinleff (b). The result from a fit including \MH as input parameter is shown in blue and the fit without \MH is shown in grey. The dotted lines indicate the fit result by setting the theoretical uncertainties \deltatheo to zero and the band corresponds to the full result. Also shown are the direct measurements and the SM prediction using a minimal set of parameters (black solid lines). 
\label{fig:mw_sinleff}}
\end{figure}

The $\Delta\chi^2$ profiles versus \MW and \sinleff without using the corresponding measurements are shown in Fig.~\ref{fig:mw_sinleff}. For the indirect determination of \sinleff all observables directly sensitive to \sinleff, like asymmetry parameters and the full and partial decay widths, are excluded from the fit.
Solid blue lines show the result of the fit including \MH, where the effect of the theory uncertainty is shown as blue band. The same fit, without information on \MH is shown in grey. An improvement in precision of more than a factor of two can be observed for the indirect determination of \MW and \sinleff. Also shown are the direct measurements of the aforementioned \pW mass and the LEP/SLD average of the effective weak mixing angle $\sinleff=0.23153
 \pm 0.00016$~\cite{LEPEWWG:2005ema}, which show good agreement with the obtained values. 
   
The fit value obtained for \MW is
 \beqn
  M_W &=& (80.3593 
          \pm 0.0056_{m_t} \pm 0.0026_{M_Z} \pm 0.0018_{\Delta\alpha_{\rm had}} \nonumber \\ 
      & & \phantom{(80.3593} 
          \pm  0.0017_{\as} \pm 0.0002_{M_H} \pm 0.0040_{\rm theo})\;{\rm GeV}\,, \nonumber \\[0.2cm]
      &=& (80.359 \pm 0.011_{\rm tot}) \;{\rm GeV} \,,
\label{eq:mw}
\eeqn
which exceeds the experimental world average in precision. The different uncertainty contributions originate from the uncertainties in the input values of the fit. 
The dominant uncertainty is due to the top quark mass, followed by the theory uncertainty of $4\:\mev$.
Due to the weak, logarithmic dependence on \MH the contribution from the uncertainty on the Higgs mass is very small compared to the other sources of uncertainty. 
The deviation between the value of \MW obtained from the fit and the direct measurement is not significant with the current precision ($1.2\sigma$). 
However, improvements in the determination of \mt as well as reduced theoretical uncertainties from higher-order calculations and a more precise direct determination of \MW~-- as expected from the analyses of the full dataset recorded by the Tevatron experiments -- will reduce the uncertainties significantly.

The indirect determination of \sinleff gives
\beqn
  \sinleff &=& 0.231496 
                 \pm 0.000030_{m_t} \pm 0.000015_{M_Z} \pm 0.000035_{\Delta\alpha_{\rm had}} \nonumber \\
           & & \phantom{0.231496}
                 \pm 0.000010_{\as} \pm 0.000002_{M_H} \pm 0.000047_{\rm theo} \,, \nonumber \\[0.2cm]
      &=& 0.23150 \pm 0.00010_{\rm tot} \;,
\label{eq:sin2t}
\eeqn
which is compatible and more precise than the average of the LEP/SLD measurements. The total uncertainty is dominated by that from the measurements of $\Delta\alpha_{\rm had}$ and $m_t$. The contribution from the uncertainty in $M_H$ is again very small.

The measurement of \MH allows for a first time to predict SM observables with a minimal set of parameters. A fit using only this minimal set of input measurements (here chosen to be \MH, $\asZ$, the fermion masses and \MZ, \GF and $\dalphaHadMZ$ for the electroweak sector) is shown by the solid black lines in Fig.~\ref{fig:mw_sinleff}. The agreement in central value and precision of these results with those from Eq.~(\ref{eq:mw}) and ({\ref{eq:sin2t}) illustrates the marginal additional information provided by the other observables once \MH is known.
  
An important consistency test of the SM is the simultaneous, indirect determination of \mt and \MW. This is particularly interesting since contributions from new physics may lead to discrepancies between the measured values and the predictions in the \mt--\MW plane. A scan of the $\Delta\chi^2$ profile is shown in Fig.~\ref{fig:mw_vs_mt} for the scenario including \MH in the fit (blue) and without \MH (grey). 
\begin{figure}
\begin{center}
\includegraphics[width=0.6\textwidth]{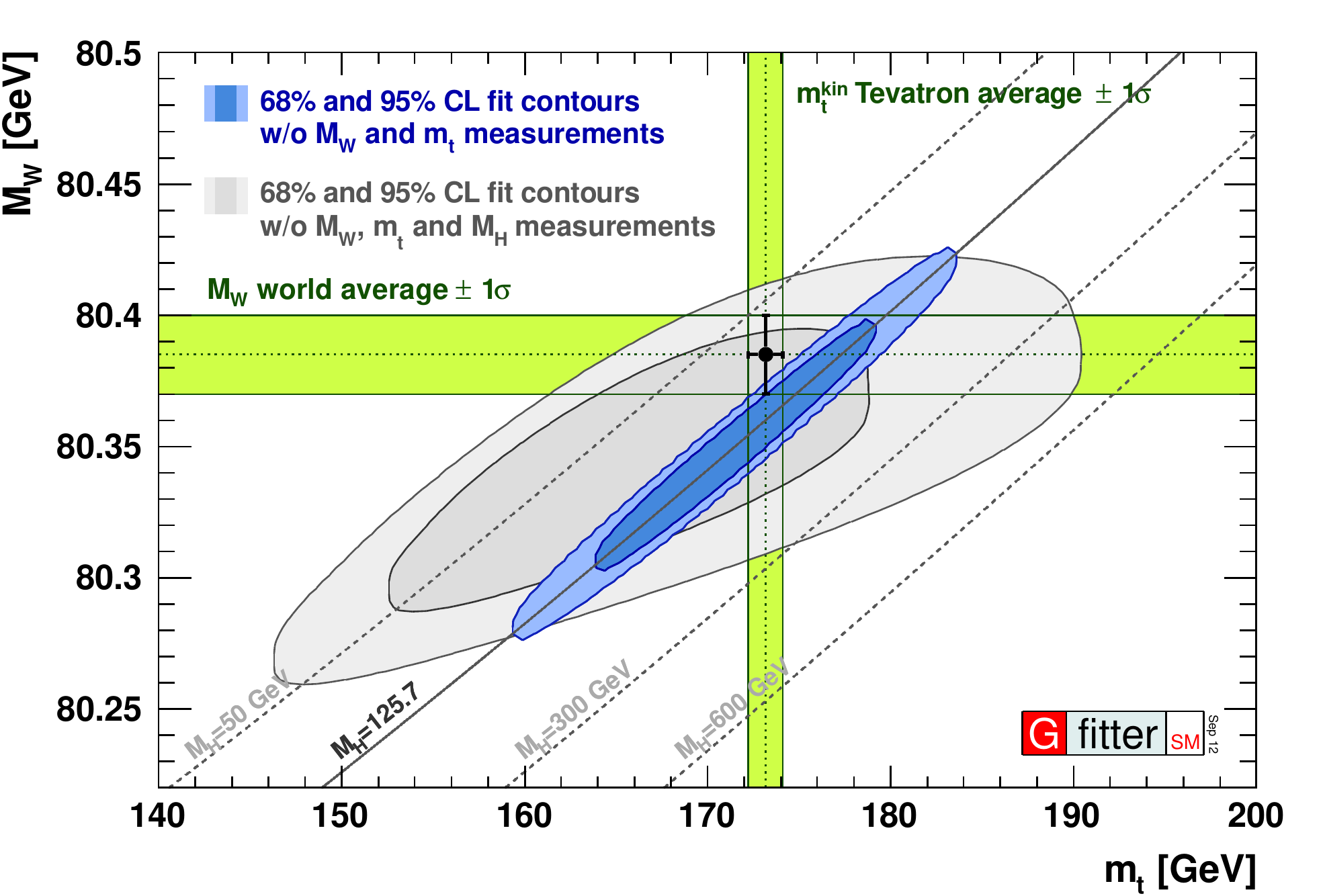}
\end{center}
\caption{
68\% and 95\% confidence level (CL) contours in the \mt--\MW plane for the fit including \MH (blue) and excluding \MH (grey). In both cases the direct measurements of \MW and \mt were excluded from the fit. The values of the direct measurements are shown as green bands with their one standard deviations. The dashed diagonal lines show the SM prediction for \MW as function of \mt for different assumptions of \MH. 
\label{fig:mw_vs_mt}}
\end{figure}
The knowledge of \MH improves the precision of the indirect determination of \MW and \mt significantly. Very good agreement between the indirect determinations of \MW and \mt and the direct measurements is observed, showing impressively the consistency of the SM and leaving little room for signs of new physics.

\section{Oblique Parameters}

If the scale of new physics (NP) is much higher than the mass of the \pW and \pZ bosons, 
beyond the SM physics appears dominantly through vacuum polarization corrections, also known as oblique corrections, 
in the calculation of EWPO. 
Their effects on the electroweak precision observables can be parametrized by three gauge 
self-energy parameters (\STU) introduced by Peskin and Takeuchi~\cite{Peskin:1990zt,Peskin:1991sw}. 
The parameter $S$ describes new physics contributions to neutral current processes at different energy scales. 
$T$ is sensitive to isospin violation and $U$ ($S+U$) is sensitive to new physics (NP) contributions to charged currents. 
$U$ is only sensitive to the $W$ mass and width, and is usually very small in NP models (often: $U = 0$).

Constraints on the \STU parameters are derived from the global fit to the electroweak precision data, 
namely from the difference between the oblique vacuum corrections determined from the experimental data 
and the corrections expected in a reference SM (defined by fixing $m_t$ and $M_H$).
The reference SM for the \STU calculation is now updated to 
$M_{H,\rm ref}=126$\,GeV and $m_{t,\rm ref}=173$\,GeV.  
With this one finds for \STU: 
\beq
  S= \SParam\:, \hspace{0.5cm}
  T= \TParam\:, \hspace{0.5cm}
  U=\UParam\:,
\eeq 
with correlation coefficients of $\STParamCor$ between $S$ and $T$, and $\SUParamCor$ ($\TUParamCor $) between $S$
and $U$ ($T$ and $U$). 
Fixing $U=0$ one obtains $S|_{U=0}= \SParamNU$ and $T|_{U=0}= \TParamNU$, with a correlation coefficient of
$\STParamCorNo$.

%
Fig.~\ref{fig:STU} shows the 68\%, 95\% and 99\% CL allowed regions in the
$(S,T)$ plane for freely varying $U$ (a) and the 
constraints found when fixing $U=0$ (b). 
For illustration purposes, also the SM prediction is shown. The \MH measurement 
reduces the allowed SM area from the grey sickle, defined by letting
\MH float within the range of $[100,1000]$\,GeV, to the narrow black strip. 
The experimental constraints on \STU can be compared now
to specific NP model predictions. 
Since the oblique parameters are found to be small and consistent with zero, 
possible NP models may only affect the electroweak observables weakly.

\begin{figure}[t]
\begin{center}
\includegraphics[width=0.5\textwidth]{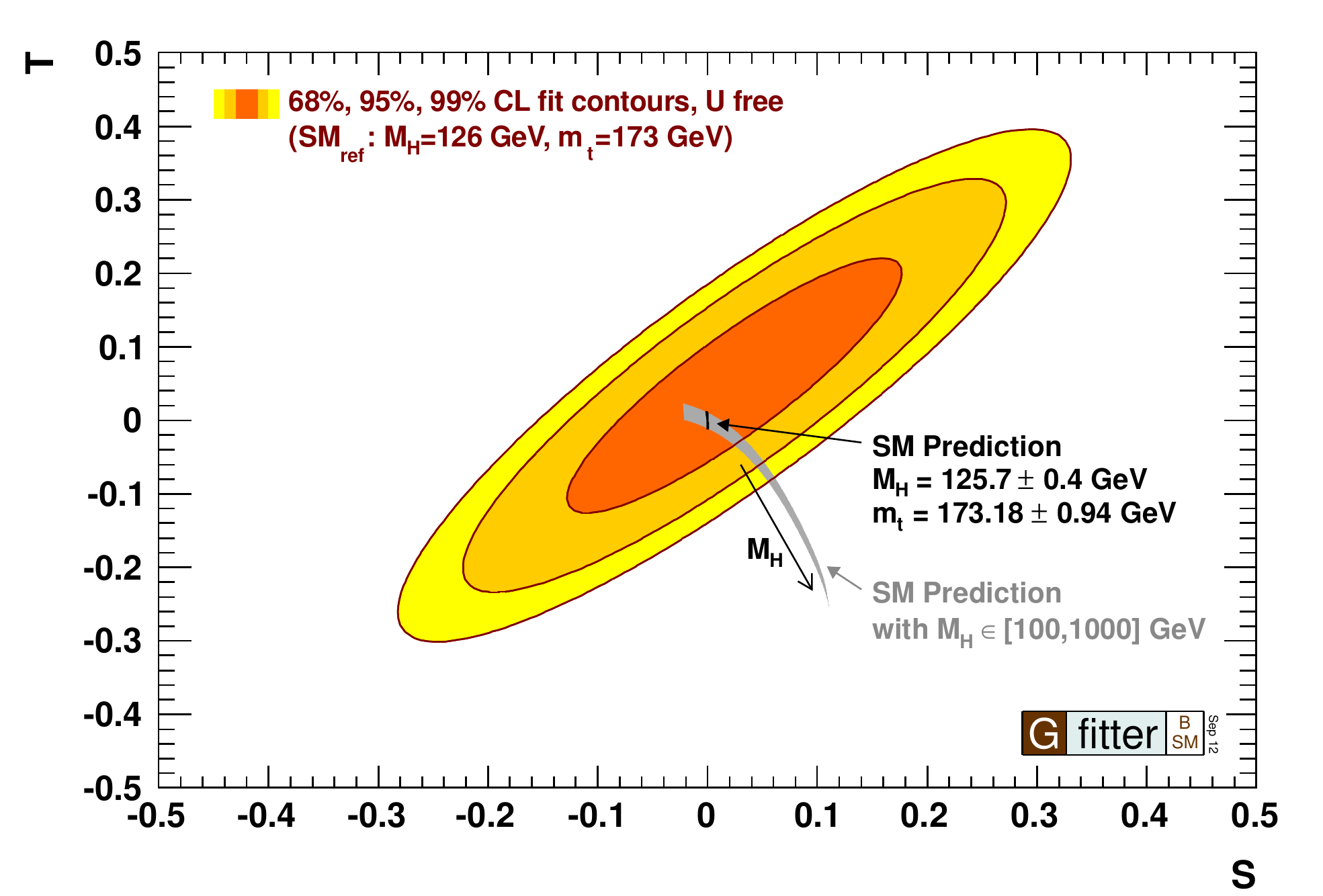} \put(-190, 114){\small (a)}
\includegraphics[width=0.5\textwidth]{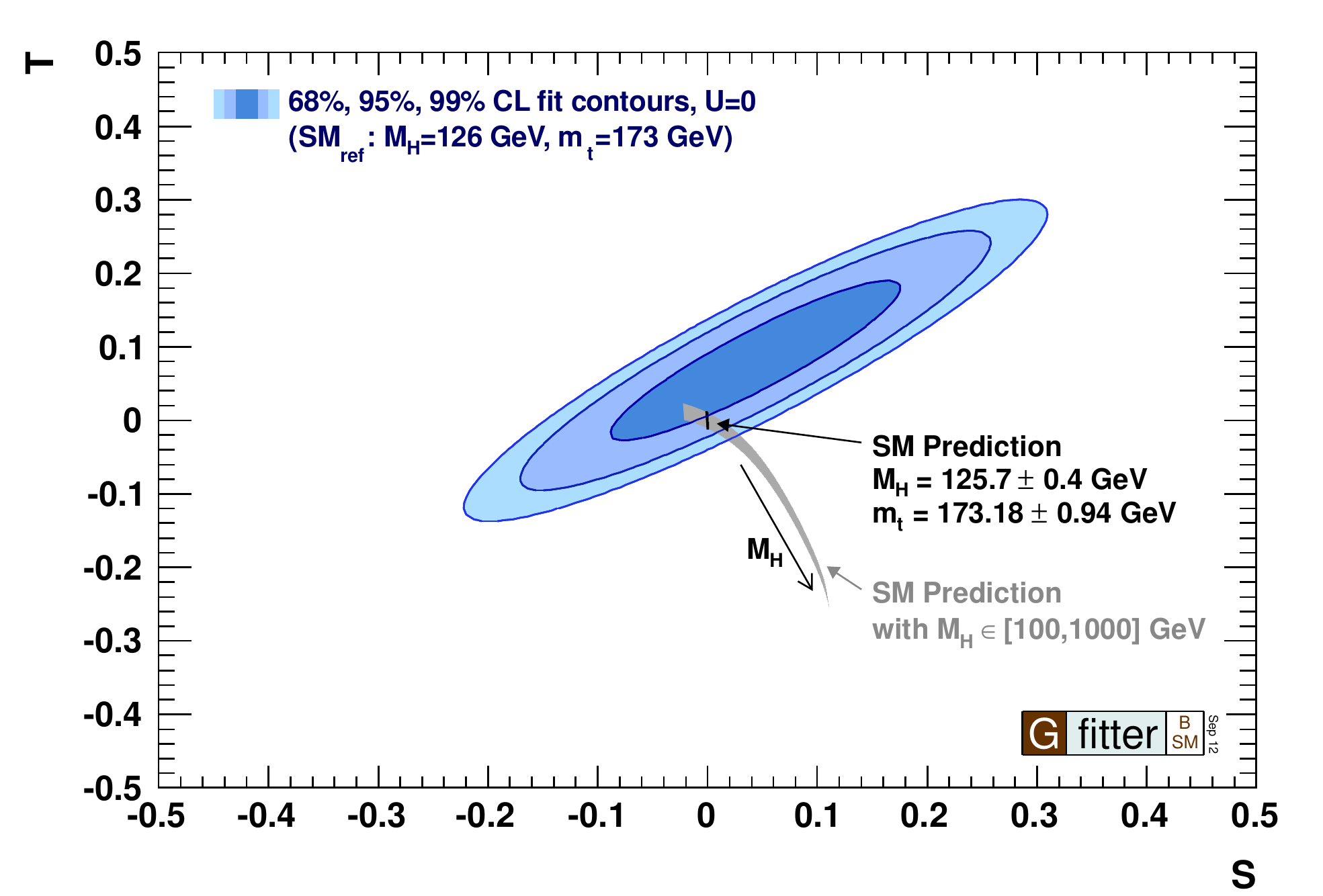} \put(-190, 114){\small (b)}
\end{center}
  \caption[]{ Experimental
    constraints on the $S$ and $T$ parameters with respect to the SM
    reference ($M_{H,\rm ref}=126\:\gev$ and $m_{t,\rm
      ref}=173\:\gev$). Shown are the 68\%, 95\% and 99\% CL allowed
    regions, where the third parameter $U$ is left unconstrained
    (a) or fixed to 0 (b). The prediction in the SM is given
    by the black (grey) area when including (excluding) the \MH
    measurements.  }
\label{fig:STU}
\end{figure}

\section{Prospects of the SM fit for a future $e^{+}e^{-}$ collider}

%
The SM leaves many questions open which can only be addressed by BSM physics, which is expected to play a role at the electroweak scale of $\mathcal{O}(1)\:\tev$. 
So far no direct signs of new physics have been observed at the LHC and also the SM shows good self-consistency at the loop-level up to very high precision.
A future $e^{+}e^{-}$ facility would help tremendously to precisely measure the production mechanisms and branching ratios of the Higgs boson. 
Furthermore, it would allow for precise measurements of the EWPO, such as \mt, \sinleff, $R^{0}_{\l}$ and \MW, to further assert the validity of the SM
through the electroweak fit. 

In the following we study the impact of expected EWPO measurements on the SM electroweak fit assuming the design parameters and predicted precisions obtained for the International Linear Collider (ILC) with the GigaZ option~\cite{TESLA:2001rg, Djouadi:2007ik,Abe:2010aa}. 
This study aims at a comparison of the accuracies of the measured and predicted 
electroweak observables. 
Two scenarios are considered. In the first scenario the central values of the input observables are 
chosen to agree with the SM prediction for a Higgs mass of $125.8\:\gev$ according to the present measurement. 
In the second scenario it is assumed that the central value of the SM prediction for \MH does not change 
and all SM parameters are chosen to agree with $\MH=94\:\gev$.

Total experimental uncertainties of $6\:\mev$ for \MW, $13\cdot10^{-5}$ for $\sinleff$, 
$4\cdot10^{-3}$ for $R^{0}_{\l}$, and $100\:\mev$ for \mt (interpreted as pole mass) are 
assumed~\cite{Djouadi:2007ik}. The exact achieved precision on the Higgs mass is irrelevant for this study.
For the hadronic contribution to the running of the QED fine structure constant 
at the $Z$ pole, $\dalphaHadMZ$, an uncertainty of
$4.7\cdot10^{-5}$ is assumed,\,\footnote{
The uncertainty on $\dalphaHadMZ$ will benefit below the charm 
threshold from the completion of BABAR analyses and the ongoing program at VEPP-2000. 
At higher energies improvements are expected from charmonium resonance data from BES-3, and a better 
knowledge of \as from the $R^{0}_{\l}$ measurement and reliable lattice QCD predictions~\cite{Davier:2012}.
} compared to the currently used uncertainty of $10\cdot10^{-5}$~\cite{Davier:2010nc}. 
The other input observables to the electroweak fit are taken to be unchanged from 
the current settings. 

The most important theoretical uncertainties in the fit are those affecting the 
$M_W$ and $\sinleff$ predictions arising from unknown higher-order corrections. 
We assume in the following that theoretical developments have led to improved uncertainties of only half the present values, $\deltatheo M_W=2\:\mev$ and $\deltatheo\sinleff=1.5 \cdot 10^{-5}$.

The indirect prediction of the Higgs mass at $126\:\gev$ achieves an uncertainty of 
$^{\,+12}_{\,-10}\:\gev$. 
Keeping the present theoretical uncertainties in the prediction of 
\MW and \sinleff would worsen the accuracy of the $M_H$ prediction 
to $^{\,+20}_{\,-17}\:\gev$, whereas neglecting theoretical uncertainties altogether 
would improve it to $\pm7\:\gev$. 
This emphasizes the importance of theoretical updates.


For \MW the prediction with an estimated uncertainty of 
$5\:\mev$ is similarly accurate as the assumed measurement, while the 
prediction of $\sinleff$ with an uncertainty of $4\cdot10^{-5}$ is three times 
less accurate than the expected experimental precision. 
The fit would therefore particularly benefit from additional 
experimental improvement in \MW. 
The accuracy of the indirect determination of the top mass is $1.2\:\gev$, which is similar to that of 
the present experimental determination. 
The fit would therefore benefit significantly from a reduction of the uncertainty on \mt to a value of about $100\:\mev$.
The measurement of $R^{0}_{\l}$ would result in a precision measurement of the strong coupling constant with
an experimental uncertainty of $0.4\%$ and a theoretical uncertainty of only $0.1\%$, which has been achieved already today 
owing to the full $\mathcal{O}(\as^4)$ calculation of the QCD Adler function~\cite{Baikov:2008jh, Baikov:2012er}.

\begin{figure}[t]
\begin{center}
\includegraphics[width=0.48\textwidth]{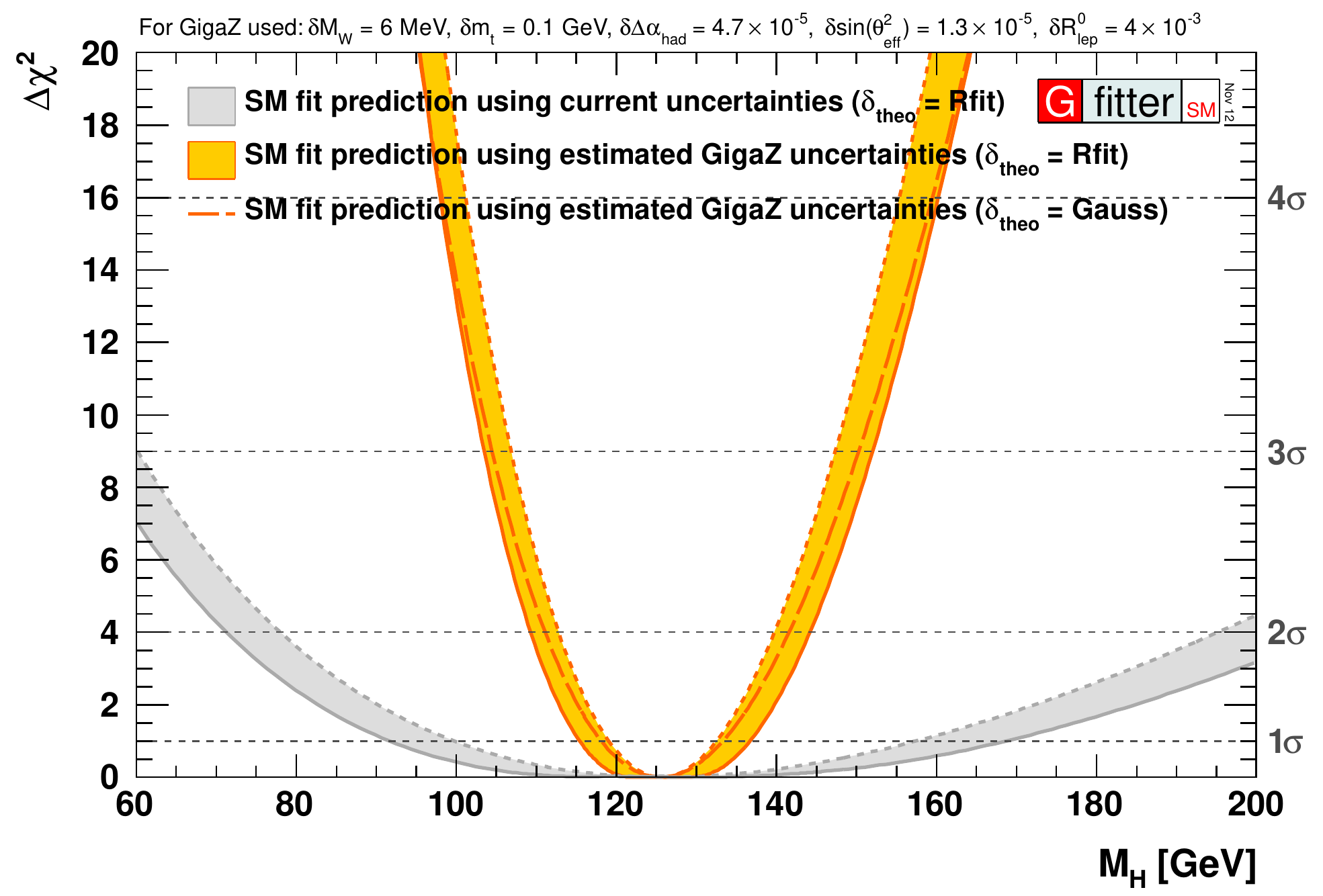} \put(-185, 97){\small (a)}
\includegraphics[width=0.48\textwidth]{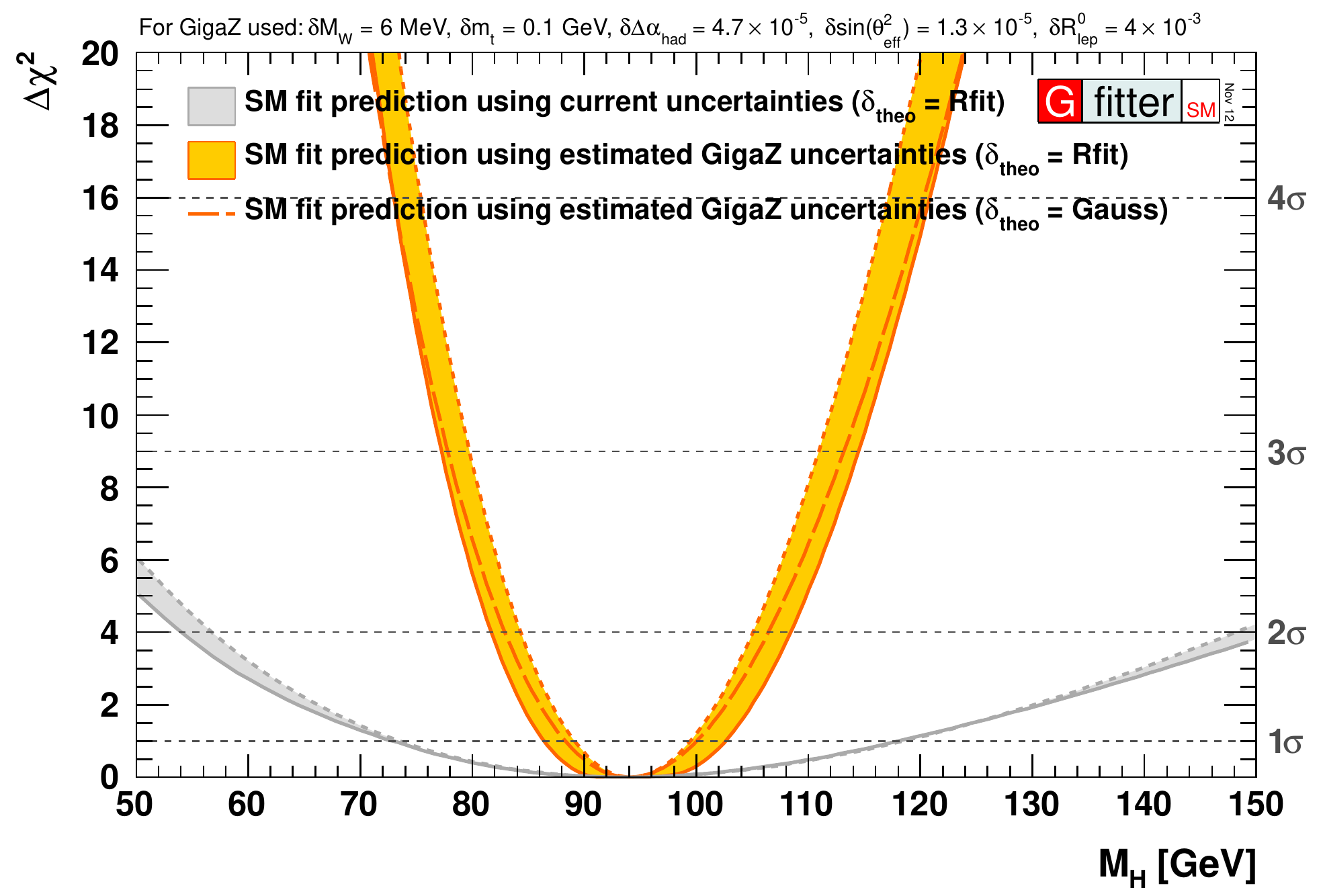} \put(-185, 97){\small (b)}
\end{center}
  \caption[]{ILC projection of the $\Delta\chi^2$ profiles as a function of the Higgs mass for 
    electroweak fits compatible with an SM Higgs boson of mass $125.8\:\gev$ (a)
    and $94\:\gev$ (b). The measured Higgs boson mass 
    is not used as input in the fit. The grey bands show the results obtained
    using present uncertainties~\cite{Baak:2012kk}, and the yellow bands 
    indicate the results for the hypothetical future scenario (a)
    and corresponding input data shifted to accommodate a $94\:\gev$ Higgs 
    boson but unchanged uncertainties (b). The thickness of the bands 
    indicates the effect from the theoretical uncertainties treated according
    to the \Rfit prescription. The long-dashed line in each plot shows the 
    curves when treating the adding the theoretical uncertainties in according to Gaussian distributed values. }
\label{fig:future_mh_scans}
\end{figure}

Profiles of $\Delta\chi^2$ as a function of the Higgs mass for present and 
future electroweak fits compatible with a SM Higgs boson of mass $125.8\:\gev$ 
and $94\:\gev$ are shown in Fig.~\ref{fig:future_mh_scans}(a) and (b), respectively. 
The measured Higgs boson mass is not used as input in these fits. 
If the experimental input data, currently predicting 
$M_H=94^{\,+25}_{\,-22}\:\gev$, were left unchanged with respect
to the present values, but had uncertainties according to the ILC expectations,
a deviation of the measured \MH exceeding $4\sigma$ could be established 
with the fit, see Fig.~\ref{fig:future_mh_scans}(b). Such a conclusion 
does not strongly depend on the treatment of the theoretical uncertainties 
(\Rfit versus Gaussian, i.e. quadratic addition) as can be seen by comparison of the solid yellow and 
the long-dashed yellow  $\Delta\chi^2$ profiles.  

\begin{figure}[t]
\begin{center}
\includegraphics[width=0.5\textwidth]{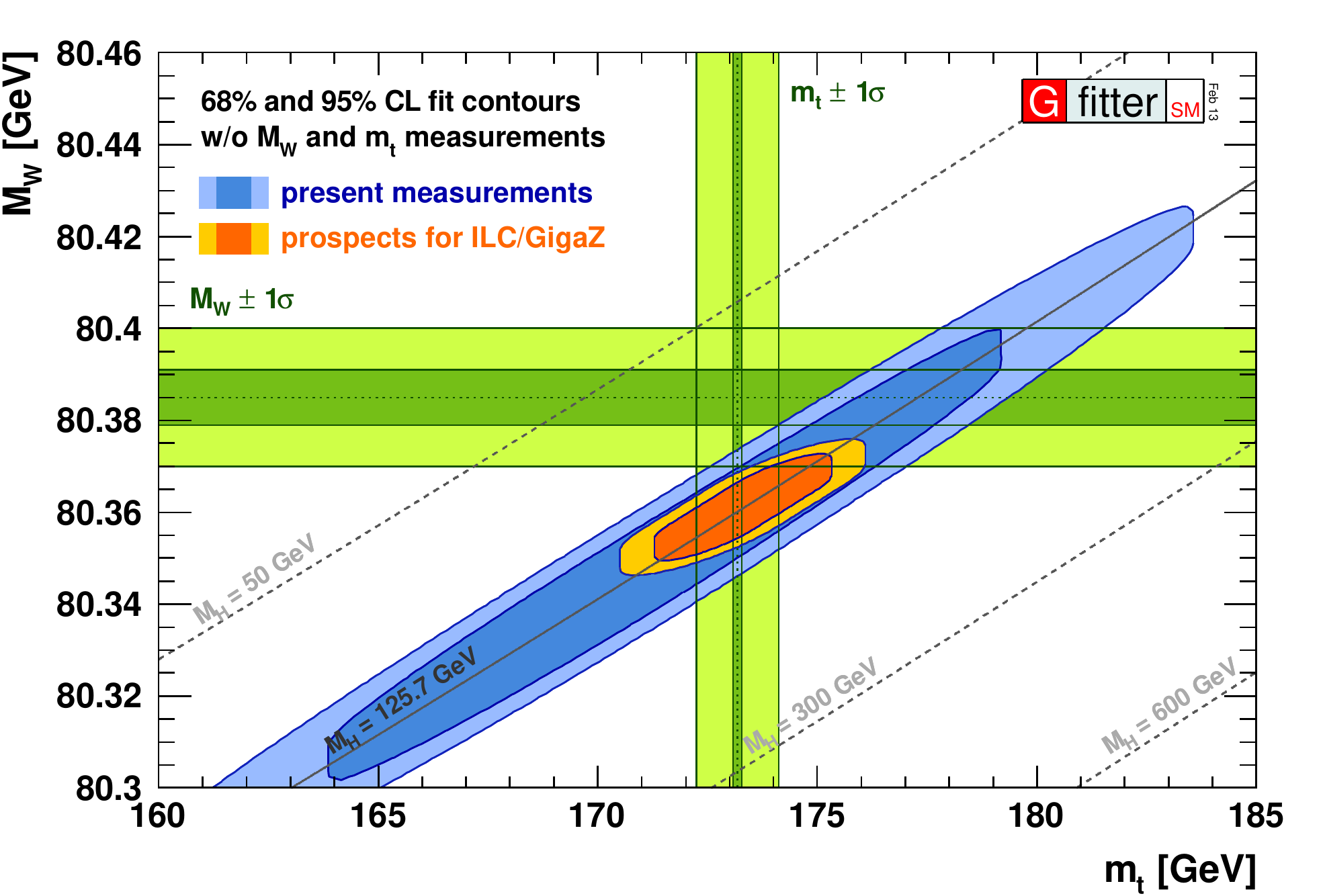} \put(-190, 63){\small (a)}
\includegraphics[width=0.5\textwidth]{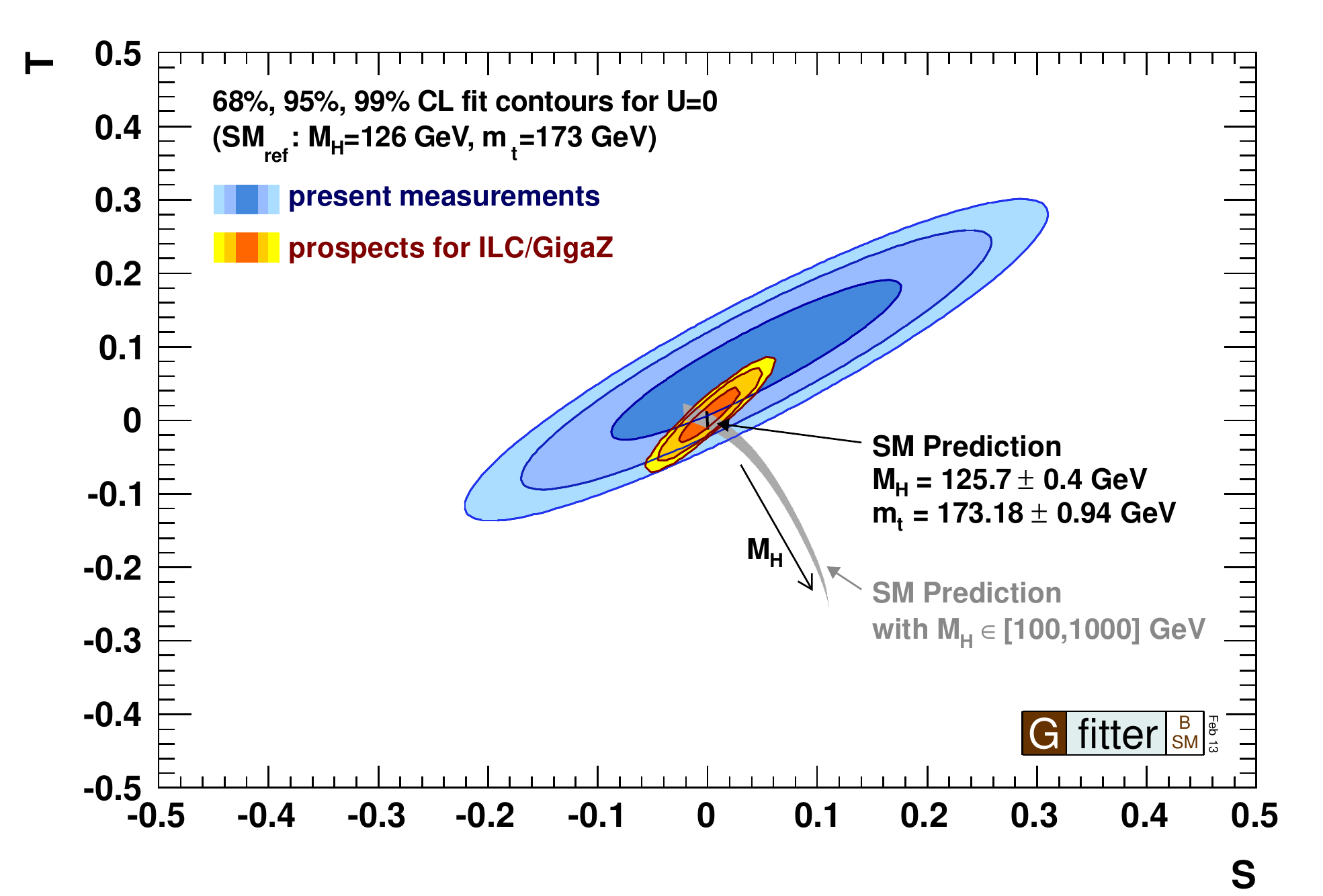} \put(-190, 97){\small (b)}
\end{center}
  \caption[]{ILC projection of the contour lines of 68\%, 95\% CL allowed regions 
    in the \mt--\MH plane. Shown are the current indirect determinations (blue) 
    and the expected precision using prospects for ILC measurements of EWPO (orange). 
    The present direct measurements together with the experimental uncertainties 
    are shown as light green bands, the prospects for the uncertainties on
    \mt and \MH are shown as dark green bands.
    Contour lines of 68\%, 95\% and 99\% CL allowed regions on the $S$ and $T$ 
    parameters for $U=0$ with respect to the SM reference 
    ($M_{H,\rm ref}=126\:\gev$ and $m_{t,\rm ref}=173\:\gev$) (b). 
    The prediction in the SM is given by the black (grey) area when including (excluding) 
    the \MH measurements.
     }
\label{fig:future_contours}
\end{figure}
Additionally to establishing a precise statement about the compatibility of the 
directly measured value of \MH and the SM prediction, the high precision 
measurements of EWPO at the ILC would significantly improve the indirect 
determination of SM parameters which has a high sensitivity to additional 
contributions of new physics on the loop-level and complements direct searches for new physics. 
Prospects for the precision of the simultaneous indirect determination of \mt and \MW
are shown in Fig.~\ref{fig:future_contours}(a) together with the present and expected precision of the \mt and \MW measurements.  
The gain in precision of the indirect measurements is about a factor of three with respect to the current determinations. 
Assuming that the central values of \mt and \MW do not change from their present values, a deviation between the SM prediction and the direct measurements would be prominently visible.

The precisely measured EWPO would also help to constrain new physics through oblique corrections. 
The expected constraints on the $S$ and $T$ parameters are shown in Fig.~\ref{fig:future_contours}(b), where 
an improvement of more than a factor of 3 seems to be possible.

\section{Conclusion}

Assuming the newly discovered particle at $\sim$126~\gev to be the Standard Model (SM) Higgs boson, 
all SM parameters entering electroweak precision observables are known. 
%
%
For the first time, the fit can over-constrain the SM at the electroweak scale and evaluate its validity.
We reported here on the most recent results from the electroweak fit~\cite{Baak:2012kk}.

The measured value of the Higgs mass agrees at 1.3$\sigma$ with the indirect prediction from the electroweak fit.
The global fit to all the electroweak precision data and the measured Higgs mass results 
in a goodness-of-fit $p$-value of 7\%. 
Only a fraction of the contribution to the ``incompatibility'' stems from the Higgs mass.
The largest deviation between the best fit result and the data is introduced 
by $A_{\rm FB}^{0,b}$ -- a known tension -- and by $R^0_b$.
%
A revisit of these two quantities 
would be very interesting, both theoretically and experimentally.

The knowledge of the Higgs mass dramatically improves the SM predictions of several key observables,
in particular of the top mass, the $W$-mass, and $\sinleff$.
The predicted uncertainties decrease by a factor of $\sim\!2.5$,
from 6.2 to 2.5~\gev, 28 to 11~\mev, and $2.3\cdot10^{-5}$ to $1.0\cdot10^{-5}$ respectively. 
Theoretical uncertainties due to unknown higher-order electroweak and QCD corrections 
contribute approximately half of the uncertainties in the \MW and $\sinleff$ predictions.

The observed agreement of these quantities between the indirect determinations and measurements demonstrates the impressive consistency of the SM.
The improved accuracy of the indirect determination of \MW sets a benchmark for new direct measurements.

Updated values for the oblique parameters \STU have been presented, using the measured value of the Higgs mass as a reference,
and indicate that possible new physics models may affect the electroweak observables only weakly.

Finally, the perspectives of the electroweak fit considering a future $e^{+}e^{-}$ collider running also at energies at the Z-pole
have been analyzed. 
Assuming a good control over systematic effects results in improved accuracy of the predictions for the top mass, the $W$-mass, $\sinleff$ and \asZ with a factor of three or greater.
We point out that, in order to fully exploit the experimental potential, theoretical developments are mandatory,
in particular in the accuracy of $M_W$ and $\sinleff$, requiring the calculation of higher order electroweak and QCD corrections.

\bibliography{References}{}

\end{document}